\documentclass[twoside]{article}

\usepackage{lipsum} 
\usepackage[hmarginratio=1:1,top=32mm,columnsep=20pt]{geometry}
\usepackage{multicol} 
\usepackage{float}

\usepackage{amsmath,amstext,amsfonts,amssymb} 
\usepackage[T1]{fontenc}
\usepackage[utf8]{inputenc}
\usepackage[english]{babel}
\usepackage{color,graphicx,graphics}
\usepackage{subfigure}
\usepackage{epsfig}
\usepackage{url}

\newcommand{\wh}{\widehat}
\newcommand{\veps}{\varepsilon}

\begin{document}
\sloppy
\title{Investigation of asymmetry in {\it E. coli} growth rate}
\author{Bernard Delyon, Beno\^\i te de Saporta, Nathalie Krell, Lydia Robert}
\date{\today}
\maketitle

\vspace{25mm}

\begin{abstract}The data we analyze derives from the observation of numerous cells of the bacterium  {\it Escherichia coli} ({\it E. coli})  growing and dividing.
Single cells grow and divide to give birth to two daughter cells, that in turn grow and divide.
Thus, a colony of cells from a single ancestor is structured as a binary genealogical tree.
At each node the measured data is the growth rate of the bacterium.
In this paper, we study two different data sets.  
One set corresponds to small complete trees, whereas the other one corresponds to long specific sub-trees. Our aim is to compare both sets.
This paper is accessible to post graduate students and readers with advanced knowledge in statistics.

	\vspace{5mm}
\textnormal{Keywords} : Autoregressive models, Binary tree, Branching processes, Dependent data, Linear regression models, Tests 
\end{abstract}

\vspace{28mm}

\begin{multicols}{2} 
%
\section{Introduction}
\begin{figure}[H]
\centerline{\includegraphics[width=.70\linewidth]{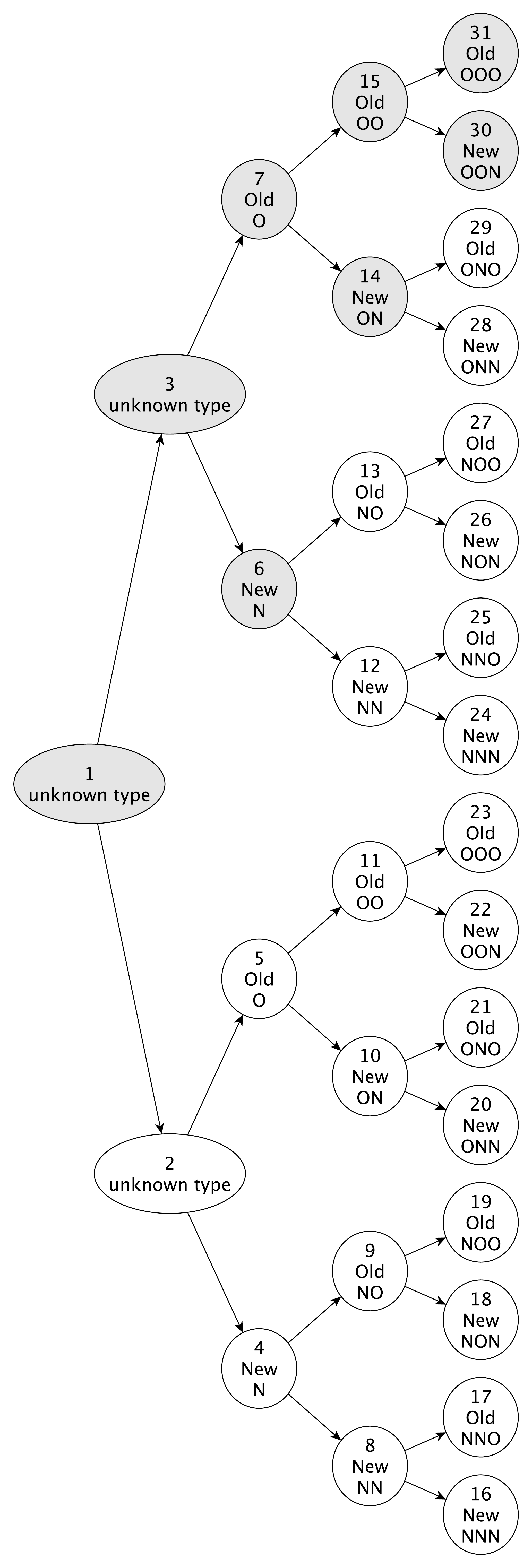}}
\caption{Cell division binary tree with the type of each cell}
\label{fig:treetype}
\end{figure}

In this paper, we study two different data sets structured as binary genealogical trees. For the statistician, this special structure is hard to take into account rigorously because of the intricate dependence structure within a tree. The data sets come from two different biological experiments. One set corresponds to small complete trees, whereas the other one corresponds to long specific sub-trees. Our aim is to compare both sets, which is especially complicated as they have very different tree structures.
\newline The underlying biological problem concerns  the growth of the bacterium {\it Escherichia coli} ({\it E. coli}).
{\it E. coli} is a rod-shaped bacterium with constant width and elongating length, hence its length (or size) is representative of its biomass or volume. Starting from size $x$ at birth, the bacterium size grows exponentially fast with time at constant rate until its division. 
More specifically, if $T$ is the age of the bacterium at division, 
there exists a constant $\tau$, which will be called the \emph{growth rate}, such that the size of the bacterium at time  $0\leq t\leq T$ equals $xe^{\tau  t}$. {\it E. coli} reproduces by binary fission, the mother cell giving birth to two virtually identical daughter cells. Because of this mode of reproduction, the observation of single cells growing and dividing for several generations produces data structured as binary genealogical trees.
Single cells growth rate within such a genealogical binary tree is the variable of interest throughout this study. 
\newline From the statistical point of view, the main difficulty in treating such data is the dependence structure as a (possibly incomplete) binary tree.
From the biological point of view, the main questions of interest are the following.
Do sister cells, that are genetically identical, have the same growth rate? Is there a memory of the growth rate between mother and daughter cells? Does it also involve the grand-mother or higher ancestors? How can it be modeled?
\newline Although two sister cells are clones with identical genetic material, asymmetry in {\it E. coli} 
division makes sense biologically. {\it E. coli} grows and reproduces by dividing roughly at its middle. Each cell has thus a new \emph{pole} (created at the division of its mother) and an old one (one of the two original poles of its mother), see Figure 1\footnote{Available at: \url{http://journals.plos.org/plosbiology/article/figure/image?size=medium&id=info:doi/10.1371/journal.pbio.0030045.g001}} in \cite{stewart}. The cell that inherits the old pole of its mother is called the \emph{old pole} cell, the other one is called the \emph{new pole} cell. It is suspected that both cells inherit different material or material of different quality from their mother cell. Therefore, each cell has a \emph{type}: old pole (O) or new pole (N) cell.
\newline On experimental data, one usually does not know the type of the original cell and its two daughters at the root of the genealogy, but from generation $2$ on, the type of each cell is known. For further generations, one can associate to one cell not only its type, but also the sequence of types of its ancestors, see Figure~\ref{fig:treetype}. The original ancestor is labelled $1$ and the two daughters of cell $n$ are labelled $2n$ for the new pole one and $2n+1$ for the old pole one. Therefore, even-labelled cells are type N and odd-labelled cells are type O and the whole sequence of types of their ancestors can be retrieved from the decomposition of their label in base 2 (with $0$ coding for N and $1$ coding for O). For instance, cell number $19$ is type NOO which means, it is type O, its mother is type O and its grand-mother is type N.
An interesting question is thus to find out whether the respective growth rate of sister cells are statistically different or not, and whether cells that have accumulated old poles along the divisions have a slower growth rate.
The starting point of the present work is that the latter questions have seemingly opposite answers in the biological literature: in \cite{stewart}, the growth rate of older cells is significantly slowed down, whereas in \cite{lydia} it is stable. We provide the data sets from both of these papers, and our aim is to conduct a new statistical study of both data sets to investigate the behavior of the growth rate of {\it E. coli} and try to decide whether both experiments yield contradictory results or not.
\newline This paper is organized as follows. In Section~\ref{sec:data sets}, we describe in details the two data sets from \cite{stewart} and \cite{lydia} and explain which statistical investigations have been conducted on each of them in papers from the literature. In Section~\ref{sec:us} we give the results of the new statistical experiments we conducted on these data sets. 
We present our conclusion in Section~\ref{sec:ccl}.
\section{Two tree-structured data sets}
\label{sec:data sets}
We first describe the data sets from \cite{stewart} and \cite{lydia} and relevant literature.
\subsection{Data set from Stewart et al.}
\label{sec:data stewart}
\begin{figure}[H]
\centering
\includegraphics[trim={0ex 0ex 0ex 12ex},width=\linewidth,clip]{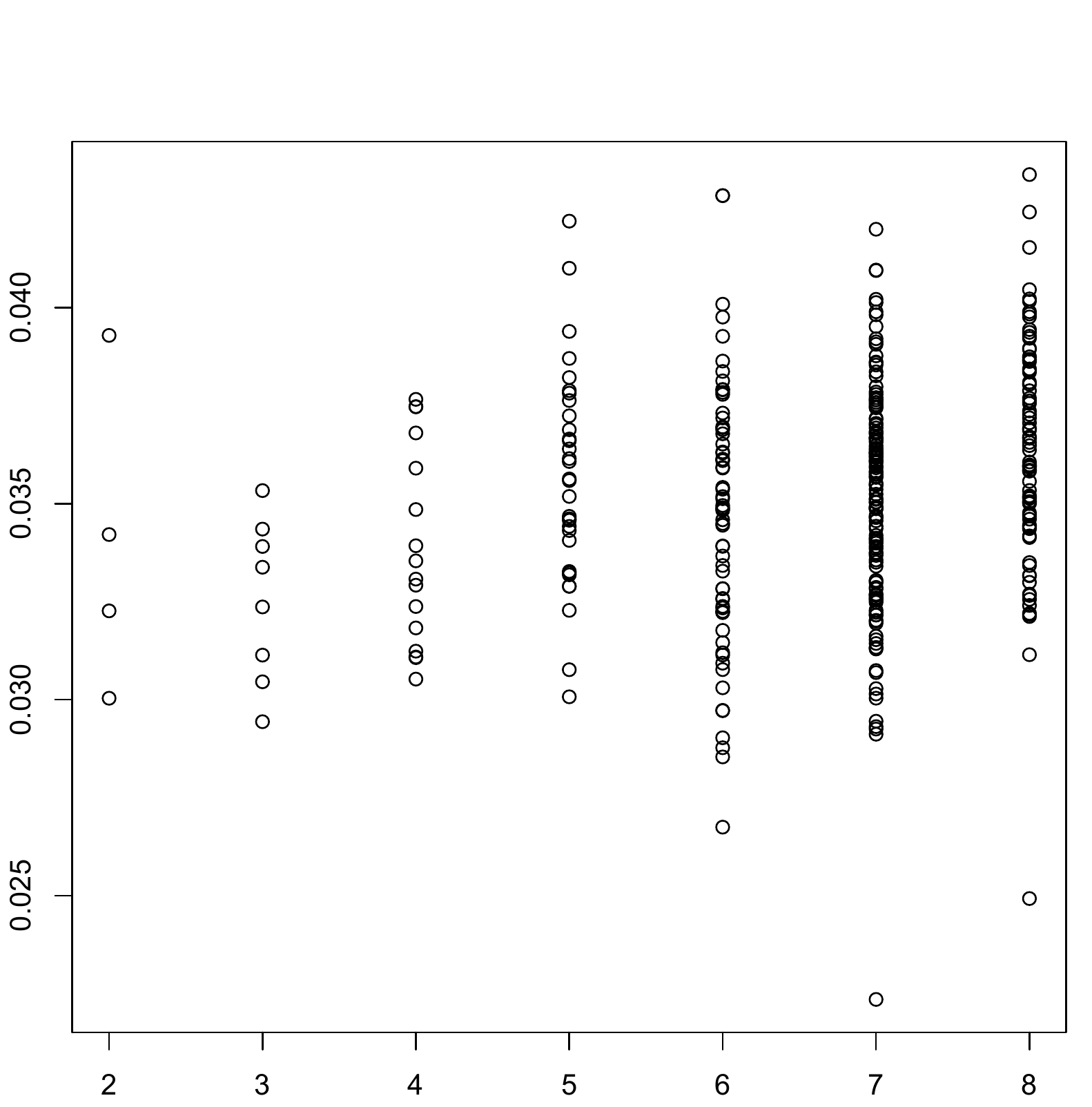}\\
\caption{Cell growth rates sorted by generation for Tree 1 in Stewart data set.}
\label{fig:tree1stewart}
\end{figure}

The first data set comes from \cite{stewart}. The authors followed the growth of 94 microcolonies of {\it E. coli}  cells by video-microscopy\footnote{For a sample film see: \url{http://journals.plos.org/plosbiology/article/asset?unique&id=info:doi/10.1371/journal.pbio.0030045.sv001}}. Each recording starts with a single cell (randomly selected from previous colonies) and stops after 7 to 9 generations of new cells. 
From the images, they measured the growth rate of $22732$ cells in $101$ (possibly incomplete) genealogical binary trees as shown in Figure~\ref{fig:treetype}. The type of each cell is also known from generation $2$ on, together with its complete lineage.
\newline In \cite{stewart}, the authors conclude that \emph{"the old pole is a significant marker for multiple phenotypes associated with aging, namely, decreased metabolic efficiency (reduced growth rate), reduced offspring biomass production, and an increased chance of death"}.
They studied the average genealogical tree and all pairs of sister cells from generation $8$ as if they were independent. More rigorous statistical studies, taking into account the dependencies induced by the tree structure, have been conducted in \cite{Guy05,Guy07,SGM11,SGM12,SGM14}. All those papers rely on the assumption of a tree-adapted autoregressive structure for the growth rate of daughter cells as a function of that of their mother, called Bifurcating Autoregressive model (BAR). All conclude that the asymmetry between the growth rate of sister cells is statistically significant.
\newline The data is provided in the file \verb+data_stewart.txt+. Each line corresponds to a single cell. There are 22732 observed cells in 101 trees (some films have multiple trees). The recorded values are given in Table~\ref{tab:stewart}. 
\begin{table}[H]
\begin{center}
\begin{tabular}{ll}
\hline
column&data\\
\hline
1& tree number	\\
2& cell number within tree\\
3& mother cell number\\
4& cell generation within tree\\	
5& mother cell generation\\
6& cell growth rate\\
7& mother cell growth rate	\\
8& no of consecutive old poles\\
9& no of consecutive new poles\\
10& no  cons. old poles for mother cell\\
11& no cons. new poles for mother cell\\
\hline
\end{tabular}
\caption{Recorded data for data set data\_stewart.txt.}
\label{tab:stewart}
\end{center}
\end{table}
Value $-1$ stands for not available.
For instance, line $100$ reads
\begin{eqnarray*}
\verb+1.  103.  51.  6.  5.  0.0348970 +&\\
\verb+ 0.0368848  3.  0.  2.  0.   +&
\end{eqnarray*}
which means cell $103$ from tree $1$ is in generation $6$, it has a growth rate of $0.0348970$. It is an old pole cell and inherited 3 consecutive old poles (type NNOOO). Its mother is labelled $51$ (note that $103=2\times 51 +1$), it belongs to generation $5$ ($5=6-1$), its growth rate is $0.0368848$, it is an old pole cell which inherited 2 consecutive old poles (type NNOO). The growth rates of tree $1$ sorted by generation are presented in Figure~\ref{fig:tree1stewart}.

%
%
\subsection{Data set from Wang et al.}
\label{sec:data wang}
\begin{figure}[H]
\centering
\includegraphics[width=\linewidth]{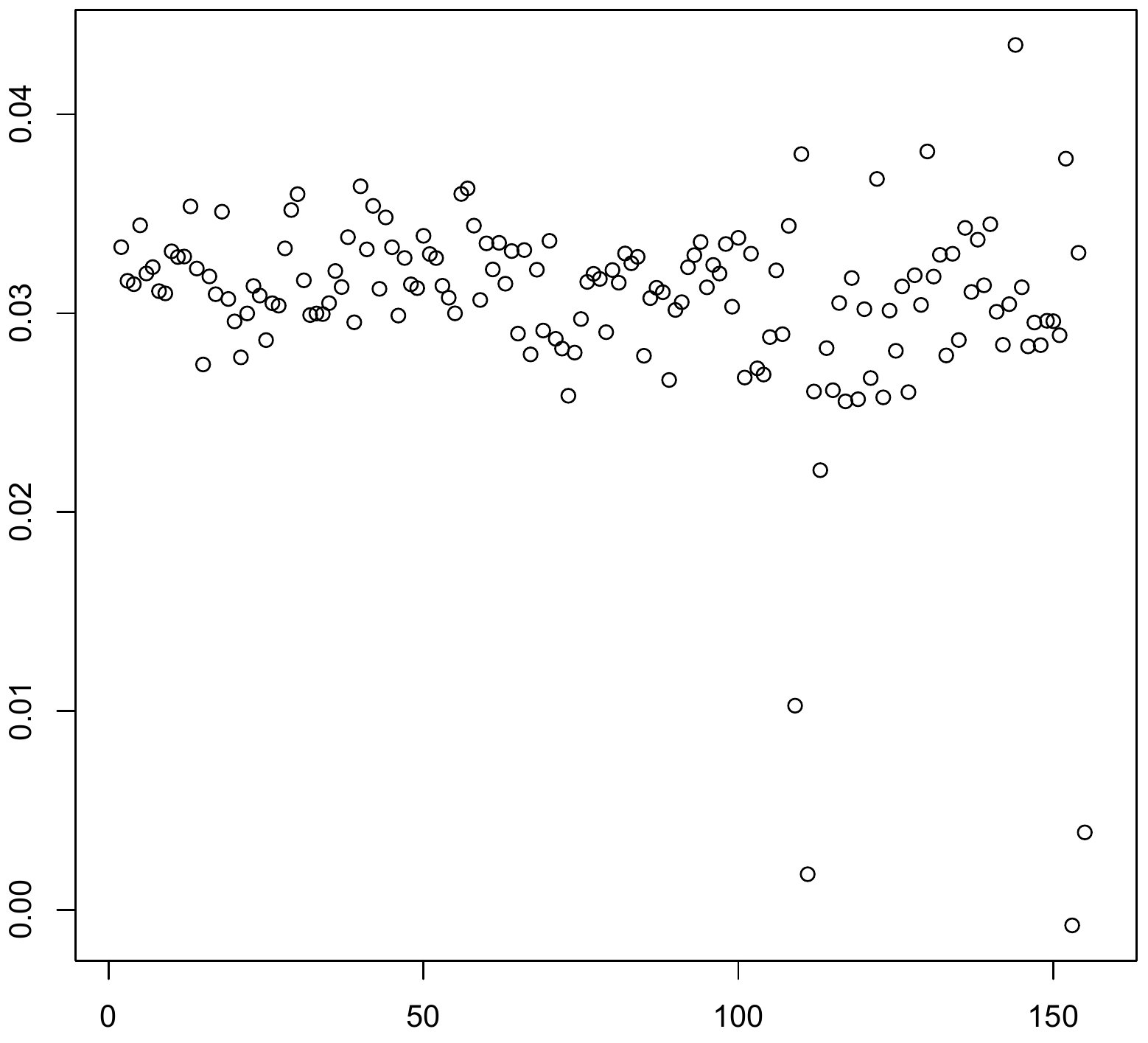}\\
\caption{Cell growth rates sorted by generation for Tree 1 in Wang data set.}
\label{fig:tree1wang}
\end{figure}

The second data set is extracted from the richer data set \cite{lydia}. The authors filmed and measured the growth and division of cells trapped in a channel, ensuring that the old pole daughter is always selected, see Figure~1\footnote{Available at: \url{http://www.ncbi.nlm.nih.gov/pmc/articles/PMC2902570/figure/F1/}. For a sample film, see: \url{http://www.ncbi.nlm.nih.gov/pmc/articles/PMC2902570/bin/NIHMS203820-supplement-03.mp4}} in \cite{lydia}. Only the cell cumulating successive old poles is observed, together with its sister. It corresponds to the grey cell sub-tree in Figure~\ref{fig:treetype}. Thus, the whole tree is not observed, but the observations can go on for a very large number of generations (up to $302$). Unlike in \cite{stewart}, the cumulated old pole cells do not exhibit a reduced growth rate but a steady state of growth. The authors conclude that they have \emph{"shown a striking constant growth rate of the mother cells of {\it E. coli} and their immediate sister cells for hundreds of generations"}.
\newline The distribution of the interdivision time of {\it E. coli} has been studied using the data set from \cite{lydia} in \cite{hof} and \cite{hof2} using a piecewise deterministic Markov process framework. In \cite{hof2}, the question is to determine which factor triggers division: the age or the size of the cell. It has been shown that the distribution of a bacterium life-time depending solely on its age does not match experimental data, while a distribution depending on size does fit the data. In \cite{hof} non-parametric statistical inference was also conducted on the experimental data to estimate the interdivision time distribution assuming division is size-dependent.
\newline To our best knowledge, this data set has not been used yet to compare the growth rate of sister cells.
\newline The data  provided in the file \verb+data_wang.txt+. Each line corresponds to a single cell. There are 45255 observed cells in 224 channels. The recorded values are given in Table~\ref{tab:wang}. 
\begin{table}[H]
\begin{center}
\begin{tabular}{ll}
\hline
column&data\\
\hline
1& tree number	\\
2& cell generation within tree\\	
3& mother cell generation\\
4& cell growth rate\\
5& mother cell growth rate	\\
6& No of consecutive old poles\\
7& No of consecutive new poles\\
8& No cons. old poles for mother cell\\
9& No cons. new poles for mother cell\\
\hline
\end{tabular}
\caption{Recorded data for data set data\_wangt.txt.}
\label{tab:wang}
\end{center}
\end{table}
We did not include the cell numbers in the trees as they grow exponentially and can be retrieved from the generation number and the type. Value $-1$ stands for not available.
For instance, line $100$ reads
\begin{eqnarray*}
&\verb+1.  50.  49.  0.0337894 +\\
&\verb+ 0.0303264  0.  1.  49.  0.   +
\end{eqnarray*}
which means cell number $2^{51}-2$ from tree $1$ is in generation $50$, it has a growth rate of $0.0337894$. It is a new pole cell. Its mother is labelled $2^{50}-1$, it belongs to generation $49$, its growth rate is $0.0303264$, it is an old pole cell which inherited 49 consecutive old poles. Note that in this data sets, old pole cells have cumulated at least, as many old poles as the rank of their generation and new pole cells always have an old pole cell mother. The growth rates of tree $1$ sorted by generation are presented in Figure~\ref{fig:tree1wang}.
%
%
\subsection{Comparison of data sets}
\label{sec:data compare}
The main difficulty for analyzing these data sets lies in the special dependence structure coming from the genealogical trees.
To take this into account, one may use the BAR model from~\cite{Guy05,Guy07,SGM11,SGM12,SGM14}. Indeed, it has been successfully applied to the first data set.
However, in the first set, one observes (almost) complete short trees, whereas on the second one, one observes very long comb-like lineages. 
This structure does not fit into the admissible observation framework of  \cite{SGM11,SGM12,SGM14} because it involves a critical Galton-Watson observation tree, where individuals of type O always have $2$ offspring, and individuals of type N always have no offspring. 
More generally, as the observed trees in both sets have a very different shape, one cannot run the same statistical procedure on both sets, making their comparison more intricate.
Last, but not least, as often with biological data, both sets are very noisy. A qualitative study may therefore be more informative than a quantitative one.
\newline The rest of this paper presents our new investigation of both data sets, the main aim being to investigate asymmetry and decide whether they lead contradictory conclusions or not.
\section{New investigation of data sets}
\label{sec:us}
%
\subsection{Preprocessing of raw data, Wang data set}
\label{sec:raw}
\begin{figure}[H]
\begin{center}
\subfigure[Tree 85]{
\includegraphics[trim={0ex 10ex 0ex 11ex},width=\linewidth,clip]{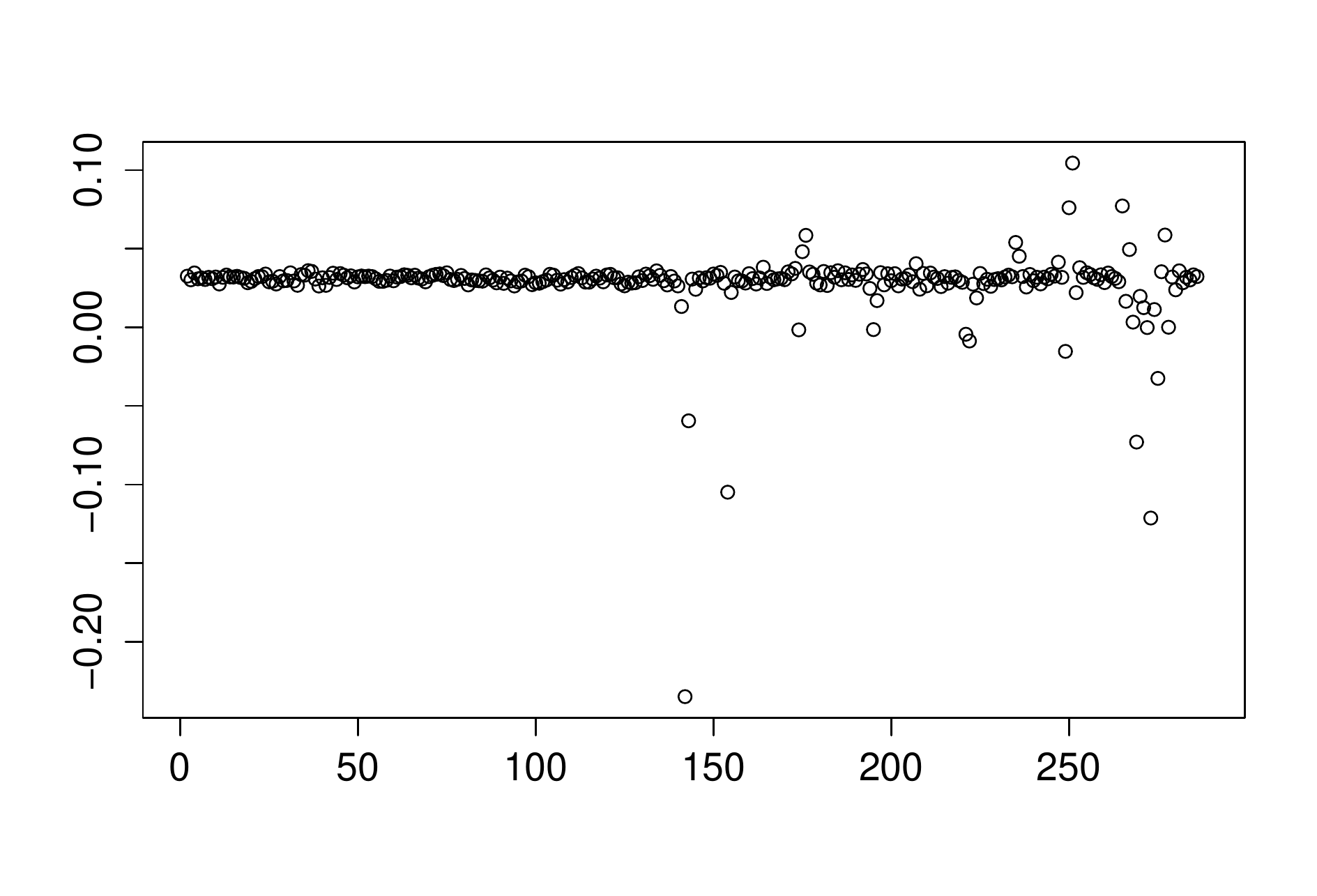}}
\subfigure[Tree 131]{
\includegraphics[trim={0ex 10ex 0ex 12ex},width=\linewidth,clip]{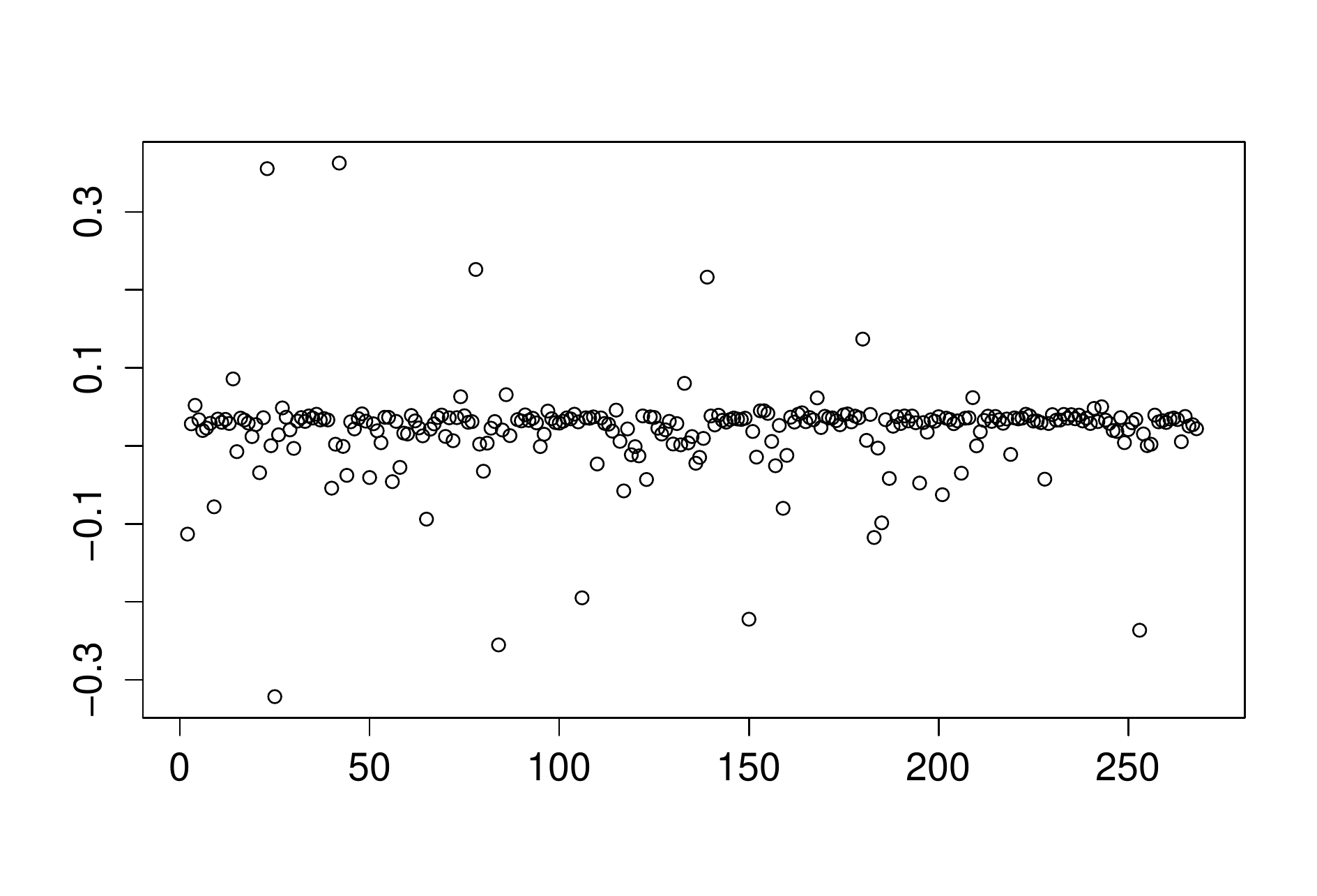}}
\subfigure[Tree 163]{
\includegraphics[trim={0ex 10ex 0ex 12ex},width=\linewidth,clip]{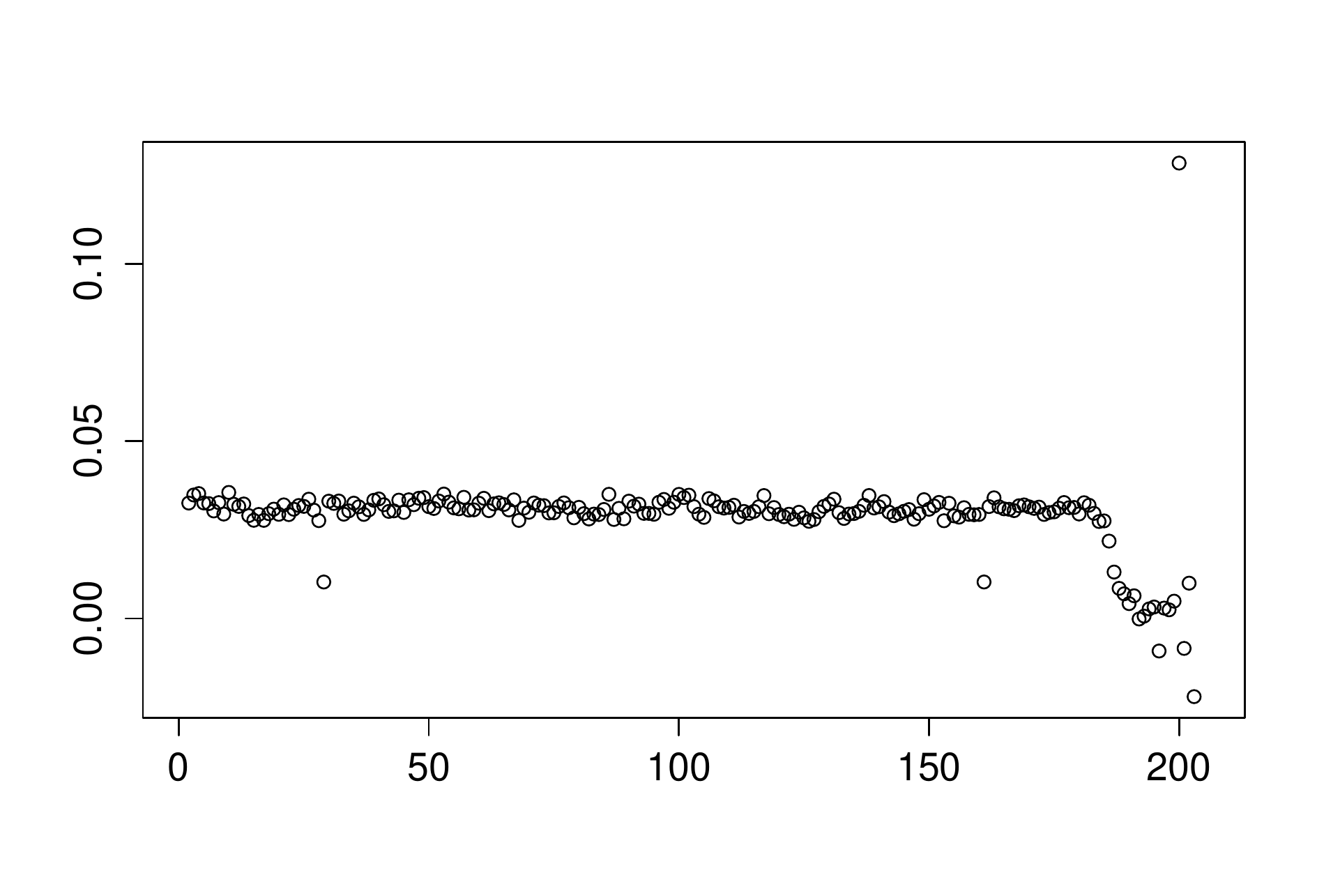}}
\subfigure[Tree 169]{
\includegraphics[trim={0ex 10ex 0ex 12ex},width=\linewidth,clip]{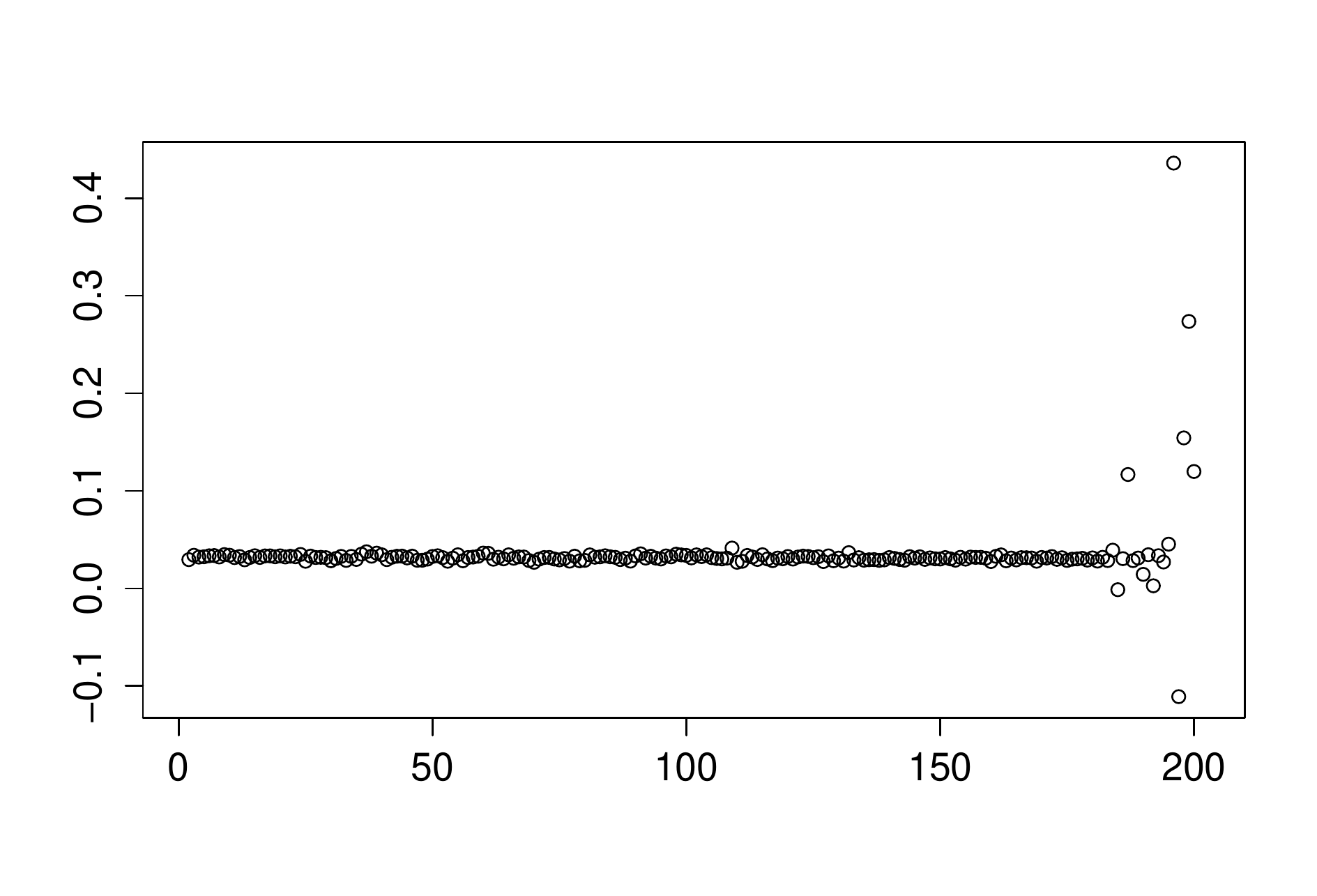}}
\caption{ Growth rate ($y$-axis) vs generation number ($x$-axis) of old pole cells, for four trees from the Wang data set.}
\label{fig:treese}
\end{center}
\end{figure}

The first difference between both data sets is that for Stewart's data, we directly received the growth rate of each cell, whereas for Wang's data, we had access to raw data of cell lengths along time and explained and computed the growth rates from an exponential regression, as presented in the introduction. When the whole life of the cell from birth to division is not observed (typically for the first and last cells in a given channel), the computation is impossible, thus we attributed the value $-1$. We did the same when some recorded length are negative. Otherwise, we provide the raw results, including possibly negative growth rates. 
\newline We first tried to work on the raw growth rates but we have quickly realized that there were too many aberrant data. Thus we developed a preprocessing based of the following observations, 
see Figure~\ref{fig:treese}:
\begin{enumerate}
\item some trees are globally aberrant (b),
\item some trees are globally good with a chaotic ending probably due to filamentation (c,d),
\item some trees are globally good with a few aberrant measures of growth rates (a).
\end{enumerate}
 This led us to  remove aberrant trees and  to mark cells with an outlying growth rate as aberrant (growth rate value set to $-1$). 
It appears that filamenting cells are automatically marked as aberrant by this procedure. 
Here is our detailed procedure.
\paragraph{Steps for preprocessing Wang data}
\begin{enumerate}
\item Remove trees smaller that 20 generations.
\item Remove aberrant trees on a criterion based on a comparison between the distribution of growth rates within this tree and the global distribution of growth rate for the whole data set:
\begin{enumerate}
\item compute robust estimates for mean $m$ and variability $\sigma$ of growth rates
over all remaining trees, using
R functions {\tt mean(.,trim=.05)} and {\tt mad(.)};
\item for each tree, compute its mean growth rate $m_t$ (usual mean), and remove the tree if 
$|m_t-m|> \sigma$.
\end{enumerate}
\item For each remaining tree:
\begin{enumerate}
\item compute the median growth rate of old pole cells, $m_O$ and of new pole cell, $m_N$;
\item mark each old pole cell whose growth rate is outside $[m_O-3*\sigma,m_O+3*\sigma]$ as  outlier;
\item mark each new pole cell whose growth rate is outside $[m_N-3*\sigma,m_N+3*\sigma]$ as  outlier.
\end{enumerate}
\end{enumerate}
\subsection{BAR model}
\label{sec:BAR}
Our first idea to compare both sets was to fit a BAR model to Wang's data set, and compare with  \cite{SGM14} where the BAR model is fitted to Stewart's data. 
It is especially appealing as the BAR model can account for a steady growth rate for the cumulative old pole lineage in the long run.

The first difficulty stems from the special comb-like structure of Wang's data trees. As explained in a previous section, it corresponds to critical Galton-Watson observation trees, thus existing results from the literature cannot be applied. However, one can readily use similar ideas as in \cite{SGM14} to propose an estimator with good convergence properties (that will not be detailed here).
\newline Let $X_{j,k}$ be the growth rate of cell number $k$ in tree number $j$, with the numeration explained in the introduction and on Figure~\ref{fig:treetype}. The asymmetric BAR model is an autoregressive model defined as follows: $X_{j,1}$ is arbitrary and for $k\geq 1$, one has
\begin{eqnarray*}
X_{j,2k}&=&a_0+b_0X_{j,k}+\veps_{j,2k},\\
X_{j,2k+1}&=&a_1+b_1X_{j,k}+\veps_{j,2k+1},
\end{eqnarray*}
where $(\veps_{j,k})$ is a noise sequence and $\theta=(a_0,b_0,a_1,b_1)$ parameters to be estimated. In order to take into account possibly missing data (in our example, they will mostly correspond to deleted aberrant values), we introduce the observation process $(\delta_{j,k})$ defined by $\delta_{j,k}=1$ if the growth rate of cell $k$ from tree number $j$ is available (i.e. not set at $-1$), $\delta_{j,k}=0$ otherwise. The mean-squares estimator of $\theta$, taking into account all the data from the $m$ trees up to generation $n$ is given by
\begin{align*}
\wh{{\theta}}_n&=\left(\begin{array}{l}
\wh{a}_{0,n}\\
\wh{b}_{0,n}\\
\wh{a}_{1,n}\\
\wh{b}_{1,n}
\end{array}\right)\\
&={S}_{n}^{-1}\sum_{j=1}^m\sum_{\ell=0}^{n-1}\left(
\begin{array}{l}
\delta_{j,2h_\ell}\delta_{j,h_\ell} X_{j,2h_\ell}\\
\delta_{j,2h_\ell}\delta_{j,h_\ell} X_{j,h_\ell} X_{j,2h_\ell}\\
\delta_{j,2h_\ell+1}\delta_{j,h_\ell} X_{j,2h_\ell+1}\\
\delta_{j,2h_\ell+1}\delta_{j,h_\ell} X_{j,h_\ell} X_{j,2h_\ell+1}
\end{array}\right)
\end{align*}
with $h_\ell =2^{\ell+1}-1$ and
where the normalizing matrix is given by
\begin{align*}
{S}_n=\left(
\begin{array}{cc}
{S}_n^0&0\\
0&{S}_n^1
\end{array}
\right),
\end{align*}
and for $i\in\{0,1\}$
\begin{align*}
{S}^i_n&=\sum_{j=1}^m\sum_{\ell=0}^{n-1}S^i_{j,\ell},\\
S^i_{j,\ell}&=\left(\!\!\!\begin{array}{cc}
\delta_{j,2h_\ell+i}\delta_{j,h_\ell}&\delta_{j,2h_\ell+i}\delta_{j,h_\ell} X_{j,h_\ell}\\
\delta_{j,2h_\ell+i}\delta_{j,h_\ell} X_{j,h_l}&\delta_{j,2h_\ell+i}\delta_{j,h_\ell} (X_{j,h_\ell})^2
\end{array}\!\!\!\right).
\end{align*}
Note that only the growth rate of cells from the comb-like subtree are taken into account, as they are the only available data in this case, i.e. cells labelled $2^{n}-1$ and $2^{n}-2$ according to the numeration described in the introduction.
\newline It can be shown with similar techniques as in  \cite{SGM14} that under mild assumptions on the noise and observation sequences, this estimator is convergent and asymptotically normally distributed. We obtain the estimation results given in Table~\ref{tab:BAR}.
\begin{table}[H]
\begin{center}
\begin{tabular}{ccc}
\hline
&Estimation&95$\%$ confidence interval\\
\hline
$\wh{a}_{0,n}$&$0.0304$&$[ 0.0200  ;  0.0410]$\\
$\wh{b}_{0,n}$&$0.0664$&$[- 0.4652  ;  0.5980]$\\
$\wh{a}_{1,n}$&$0.0281$&$[ 0.0178   ; 0.0385]$\\
$\wh{b}_{1,n}$&$0.0994$&$[- 0.3194  ;  0.5182]$\\
\hline
\end{tabular}
\caption{Estimated parameters for the BAR model, Wang data, $n=302$, $m=224$.}
\label{tab:BAR}
\end{center}
\end{table}
The estimated variance of the noise sequence is very high (about $.5$) compared to the magnitude of the data, leading to wide confidence intervals. In particular, as $0$ belongs to the confidence intervals of $\wh{b}_{0,n}$ and $\wh{b}_{1,n}$ (refer to Table \ref{tab:BAR}) one cannot assert that the autoregressive structure is relevant, and we cannot rely on this model to test the symmetry of old and new pole cells. 
\newline How to deal with the high level of noise is an important question for this data set.
We tried imputation methods for missing values due to aberrant marking,
but it appeared that this introduced a strong bias in the tests. We observed that
the analysis was very sensitive to the choice of the imputation method, thus we gave up the idea and went on working with uncorrected non aberrant data.
%
\subsection{Memory from the mother and higher ancestors, Wang data set}
\label{sec:ancestors}
\begin{figure}[H]
\begin{center}
\subfigure[]{
\includegraphics[trim={2ex 2ex 2ex 8ex},width=\linewidth]{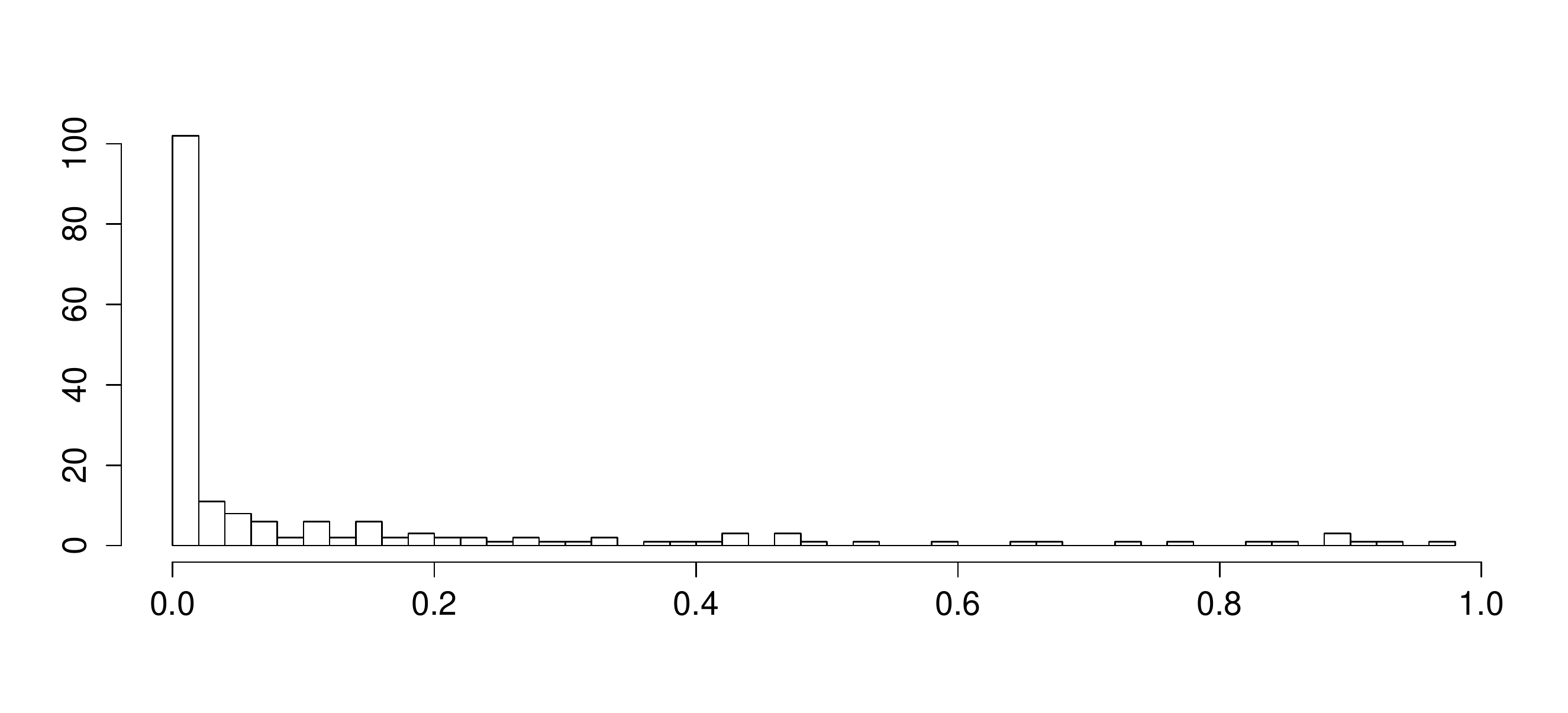}}
\subfigure[]{\includegraphics[trim={2ex 2ex 2ex 8ex},width=\linewidth]{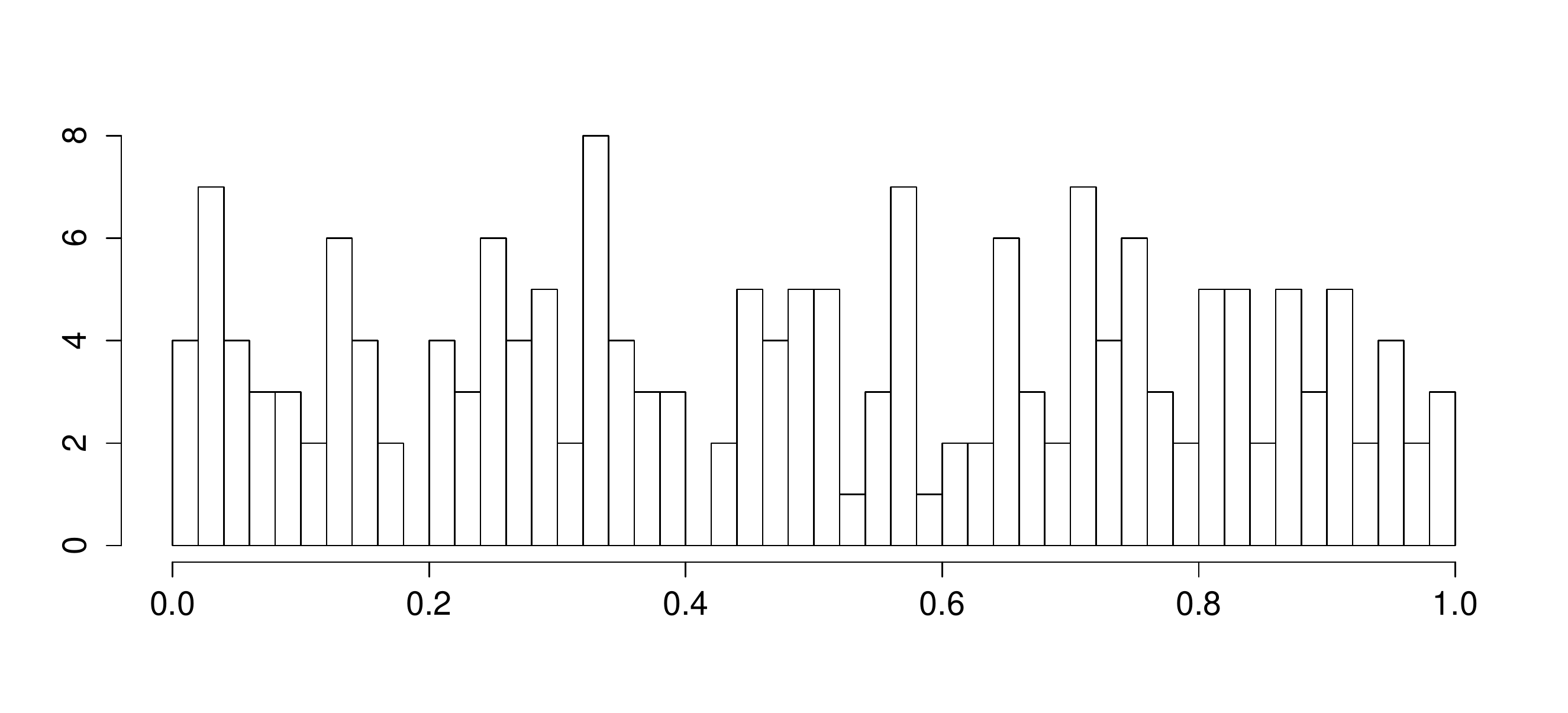}}
\end{center}
\caption{Histogram of p-value for significance of the mother coefficient $\beta_m$  (a) and for the grand mother coefficient $\beta_g$ (b), Wang data set.}
\label{fig:figmg}
\end{figure}

For each tree, we selected the old cell branch (upmost branch in Figure~\ref{fig:treetype}) and we fit an additive regression model explaining the growth rate of a cell with the  one of its mother and the one of its grand mother
 \begin{align}
 \label{reg}
 r_n=\beta_m m_n+\beta_gg_n+\beta_0 +e_n
 \end{align}
 where 
 \begin{itemize}
\item $r_n$ is the growth rate of the $n$-th generation cell ($X_{2^{n+1}-1}$ with previous notation),
\item $m_n$ is the growth rate of its mother ($X_{2^{n}-1}$),
\item $g_n$ is the growth rate of its grand mother ($X_{2^{n-1}-1}$)
\item $e_n$ the prediction error.
\end{itemize}
The triple ($\beta_0$, $\beta_m$, $\beta_g$) depends on the tree. 
The R command is {\tt lm(rate$\sim$ratemo+rategdmo)}.
Histograms of p-values for the significance of the mother coefficient $\beta_m$ (a) and for the grand mother coefficient $\beta_g$ (b) are plotted in Figure~\ref{fig:figmg}.
We conclude that the effect of the grand mother is not significant. 
The coefficient $\beta_m$ is significantly positive with a value around $0.3$.
%
%
\subsection{Comparison of old pole and new pole statistics, both sets}
\label{sec:stationarity}
\begin{figure}[H]
\centering
\subfigure[]{
\includegraphics[width=.98\linewidth]{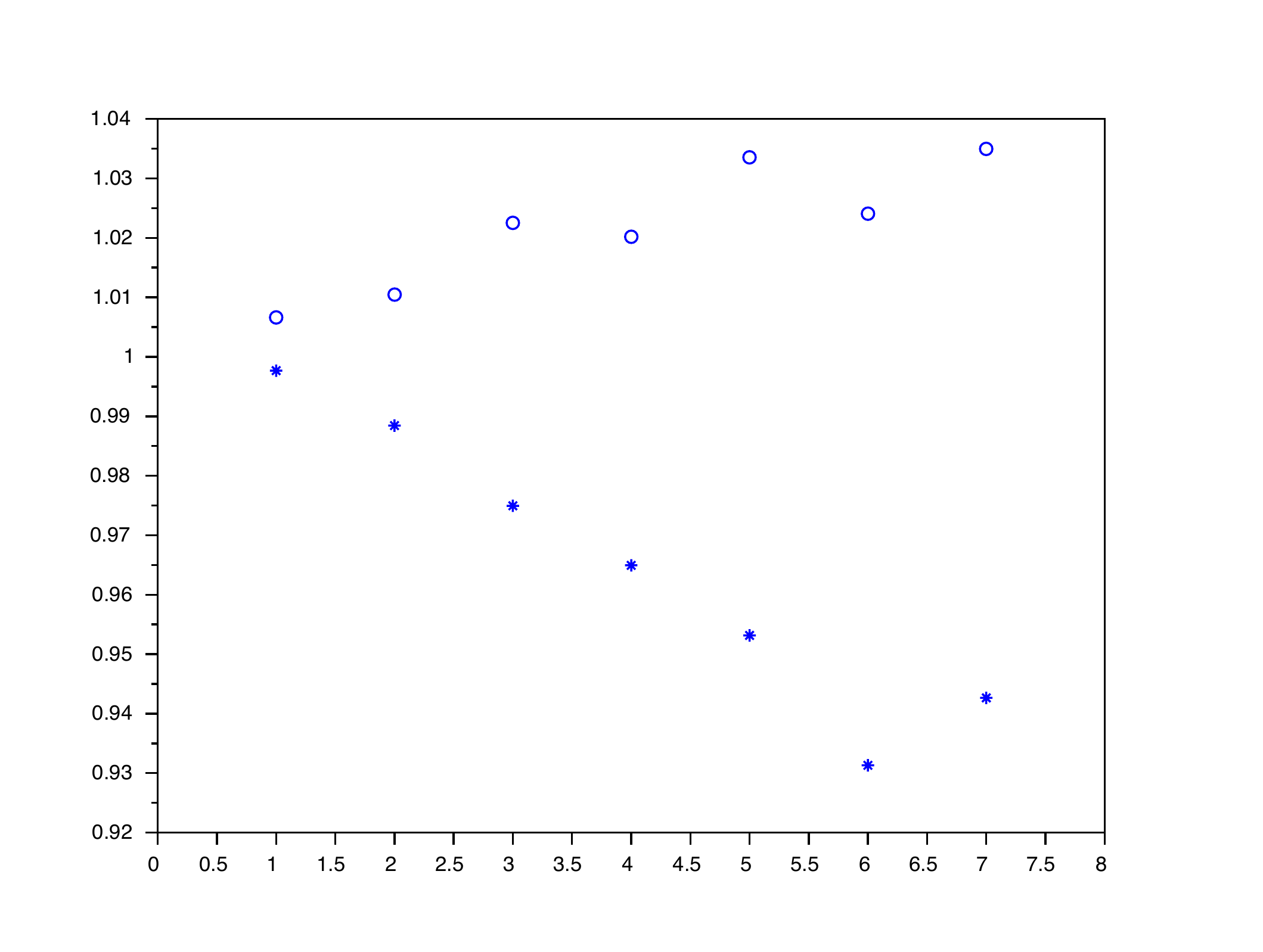}}
\subfigure[]{
\includegraphics[width=.98\linewidth]{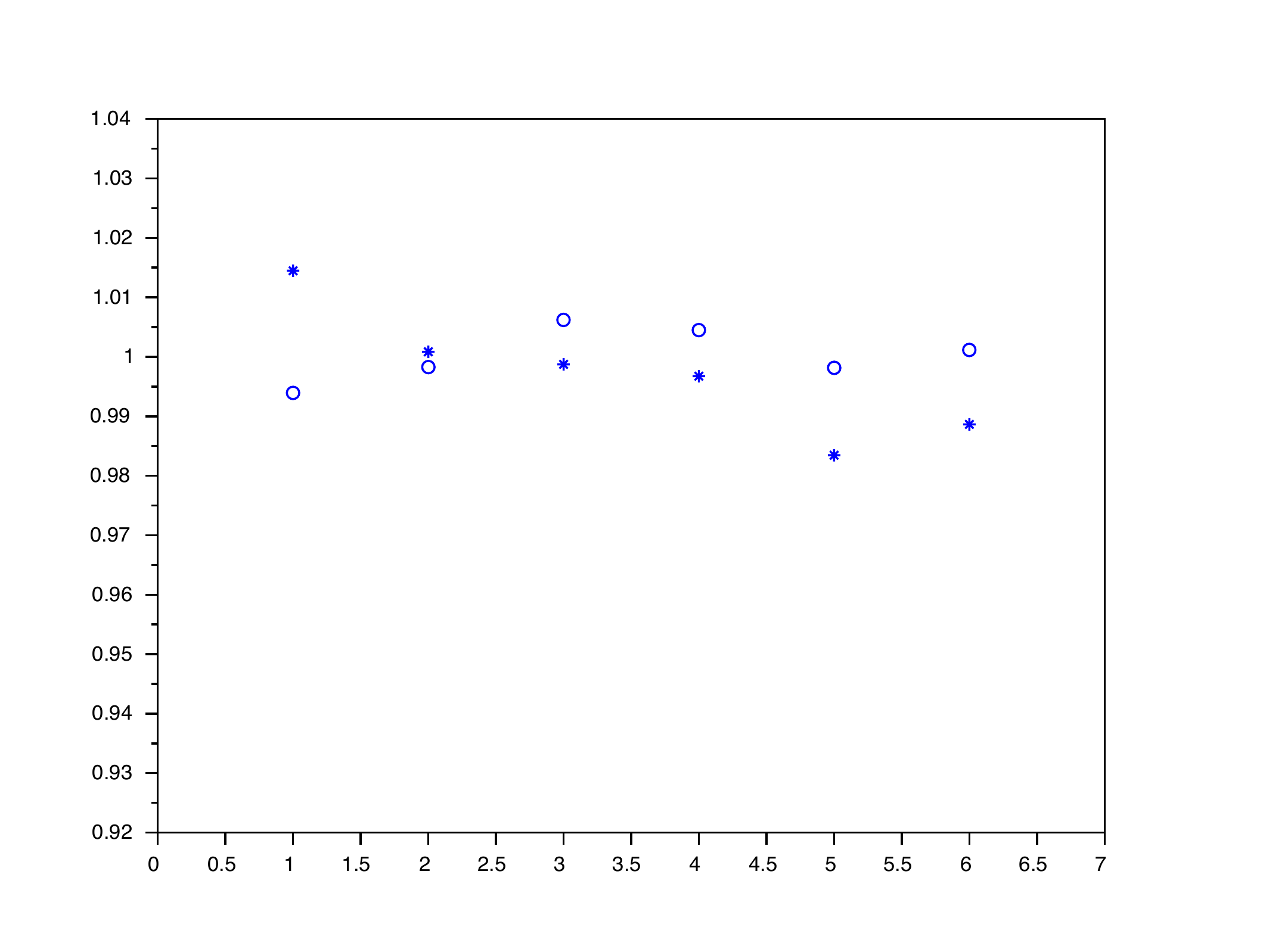}}
\caption{Mean normalized growth rate within generations and trees for cells that have cumulated (a) $n$ consecutive new poles (circles) or $n$ consecutive old poles (stars) for $1\leq n\leq 7$; (b) 1 new pole after $n$ consecutive old poles (circles), 1 old pole after $n$ consecutive new poles (stars),  for $1\leq n\leq 6$, Stewart data set.}
\label{fig:nNnO}
\end{figure}

As it is not possible to compare the BAR model for both data sets, we turned to more basic tools to compare the influence of the mother and higher ancestors on the growth rate of a given cell. 
Here again, as both data sets do not have the same structure, one cannot run the exact same experiments on both sets. Recall that asymmetry is already proved rigorously for Stewart's data.
\newline
The authors in \cite{stewart} averaged and normalized the growth rate data within each generation and each tree (combined with another indicator of distance to the edge of the microcolony) to obtain their Figure 3\footnote{Available at \url{http://journals.plos.org/plosbiology/article/figure/image?size=large&id=info:doi/10.1371/journal.pbio.0030045.g003}} showing a linear increase (respectively decrease) for the mean normalized growth rate of cells with cumulated consecutive new poles (respectively cumulated consecutive old poles). Although the lower generations contain significantly fewer individuals than higher generations and cells with an identical number of cumulated old/new poles can exist within the same genealogical tree, we used the same approach to try to find out how many new poles it requires to obtain a rejuvenated cell (with respect to its growth rate).
 \newline We averaged the growth rates of cells within the same generation of the same tree (irrespectively of the edge distance), and normalized the growth rate of each cell with the corresponding average. 
Then we computed the mean growth rate over all normalized cells that have cumulated $n$ new poles or $n$ old poles (for $1\leq n\leq 7$). 
\newline
The results are given on Figure~\ref{fig:nNnO} (a), circles correspond to cumulated new-pole cells and stars to cumulated old-pole cells. This figure corresponds to Figure 3 in \cite{stewart}. Then we compared the mean of all new-pole cells which mother cumulated $n$ old poles, and old-pole cells which mother cumulated $n$ new poles (for $1\leq n\leq 6$), see Figure~\ref{fig:nNnO} (b), circles correspond to new-pole cells with cumulated old-pole mother and stars to old-pole cells with cumulated new-pole mother. The scales of both figures are the same to make visual comparison easier.
The linear regression slope coefficients are respectively $4.4\%$ for the new pole cells and $-1.1\%$ for the old pole ones in Figure~\ref{fig:nNnO}~(a), $0.1\%$  for the new pole cells and $-0.5\%$ for the old pole ones in Figure~\ref{fig:nNnO}~(b).
\newline One can conclude that one new pole is enough to \emph{forget} an accumulation of old poles and similarly one old pole is enough to forget an accumulation of new poles. 
%
As regards Wang's data, we compared the mean growth rate of new pole and old pole cells as well as mother-daughter correlation. More specifically, we found out the following.
%
\begin{enumerate}
\item  Student test for comparison of the mean of the growth rate of old pole cells
 and of new pole cells yields a p-value $<10^{-16}$, and 
$1\%$  confidence intervals for mean growth rates are:
$[0.0309, 0.0310]$ for old pole cells and $[0.0319, 0.0320]$ for new pole cells.
\item Regarding the daughter~mother correlation, we have computed one confidence interval
 for the overall correlation between the growth rate of old pole daughters and their mothers', and another one for new pole daughters and their mothers':
1\% confidence intervals for correlation between growth rates of new pole daughters and that of their mother is $[0.085, 0.123]$, the same for old pole cells
is $[0.125, 0.160]$.
\end{enumerate}
 A significant difference thus holds for the mean as well as for the correlation with the mother cell for old pole and new pole sister cells.
\begin{figure}[H]
\begin{center}
\subfigure[New pole cells]{\includegraphics[trim={2ex 2ex 2ex 8ex},width=\linewidth]{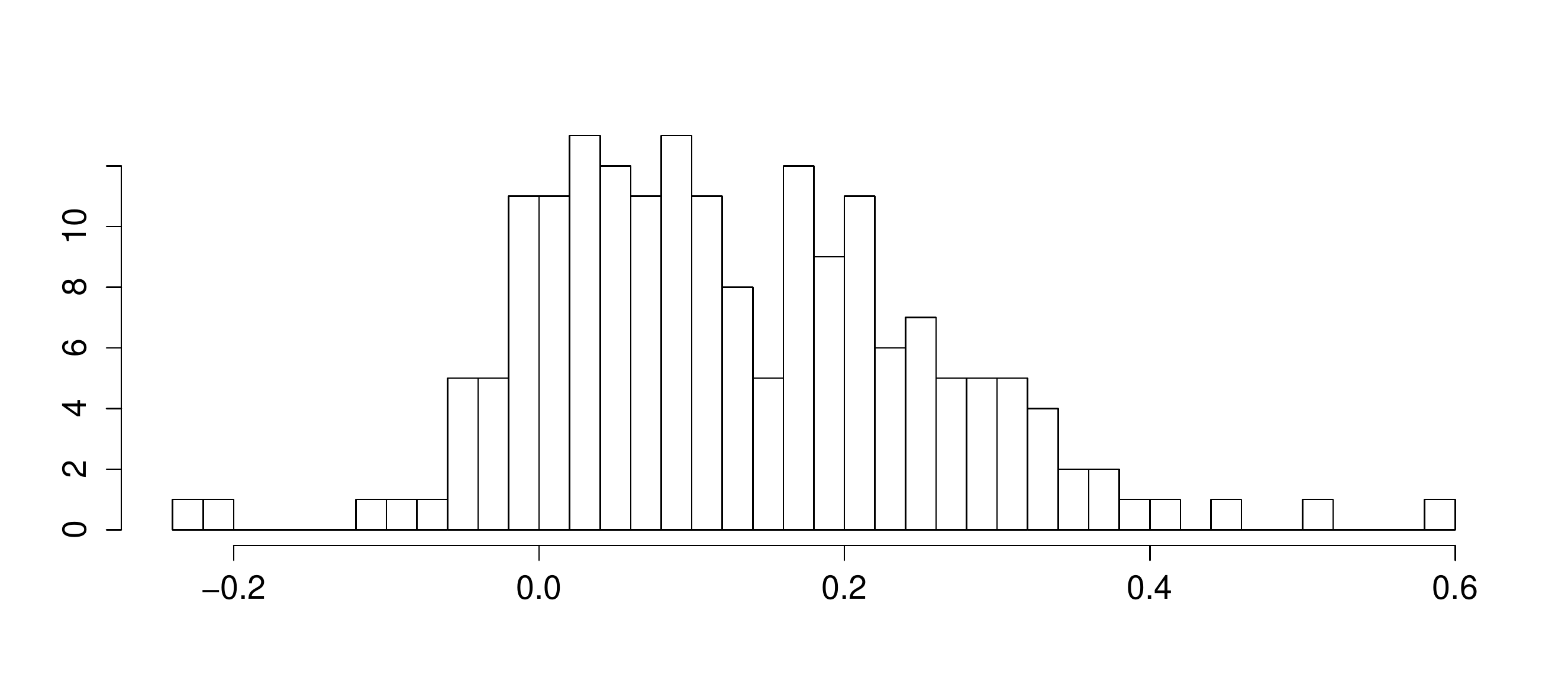}}
\subfigure[Old pole cells]{\includegraphics[trim={2ex 2ex 2ex 8ex},width=\linewidth]{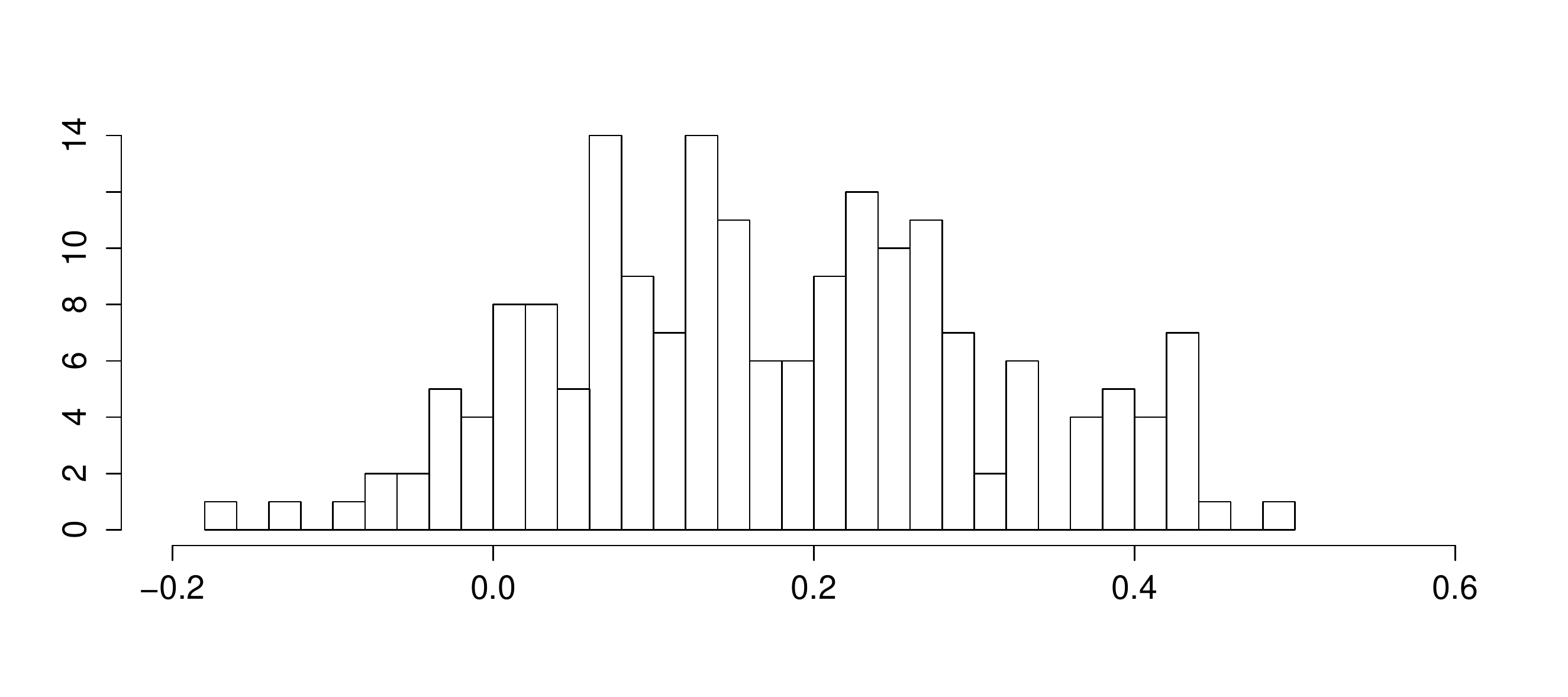}}
\caption{Histogram of regression coefficients $\beta_m$, for new poles cells (a) and for old poles cells (b), Wang data set.}\label{fgmg}
\end{center}
\end{figure}
We have also plotted, in Figure~\ref{fgmg}, a histogram of regression coefficients (w.r.t. mother's growth rate) 
in both cases, corresponding to coefficients $\beta_m$ in Equation (\ref{reg}) with $\beta_g$ set to $0$. %
The difference in not clear, but it seems that in the case of old poles,
the dispersion is smaller.
%
\subsection{Stationarity, both sets}
\label{sec:stationarity}
\begin{figure}[H]
\centering
\includegraphics[trim={0ex 0ex 0ex 0ex},width=\linewidth,height=.65\linewidth,clip]{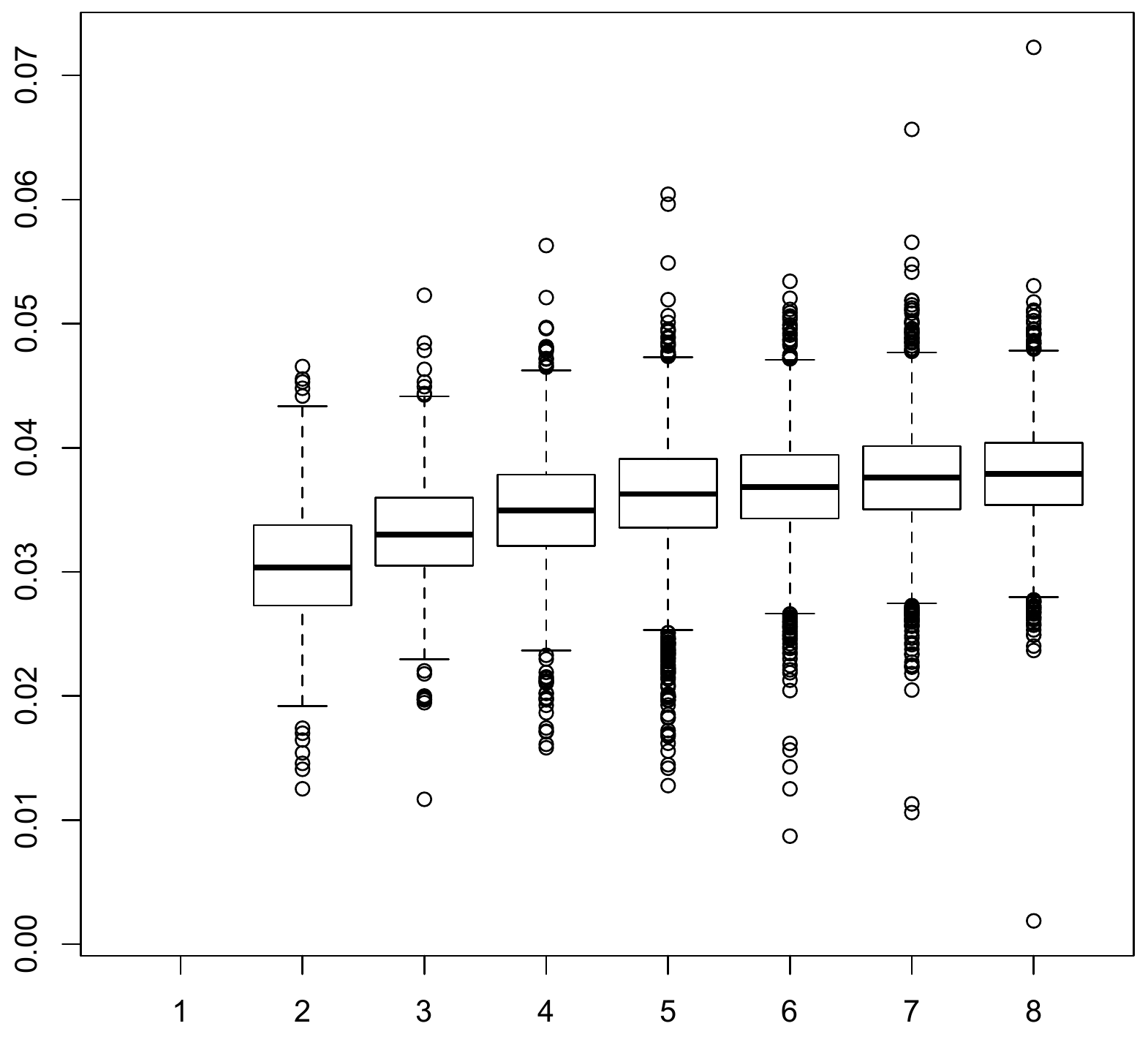}
\caption{Box plots of growth rates for cells in generations 2 to 8, Stewart data set.}
\label{fig:boxp-stewart}
\end{figure}

\begin{figure}[H]
\centering
\subfigure[Generation 2]{\includegraphics[width=.9\linewidth]{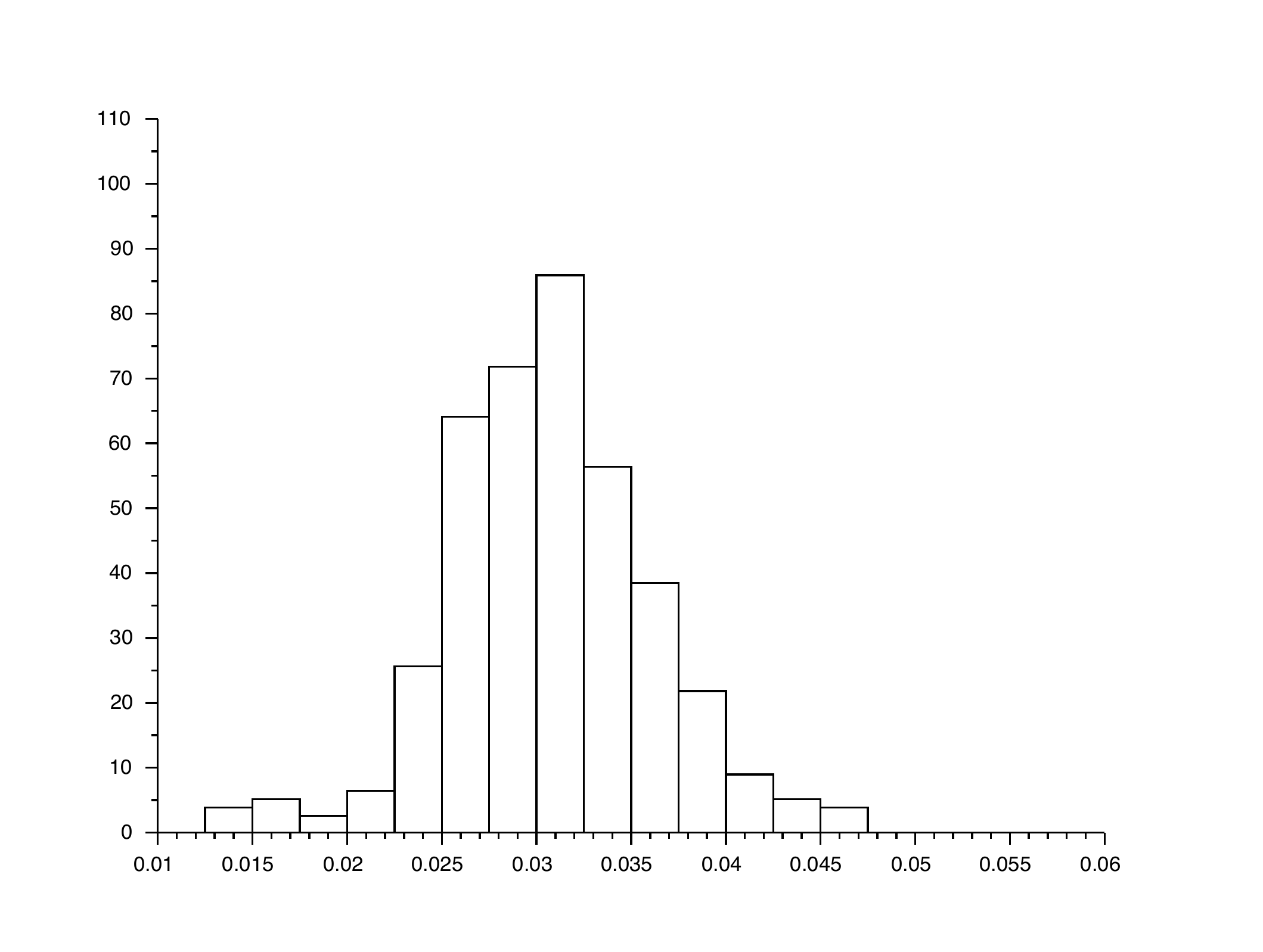}}
\subfigure[Generation 5]{\includegraphics[width=.9\linewidth]{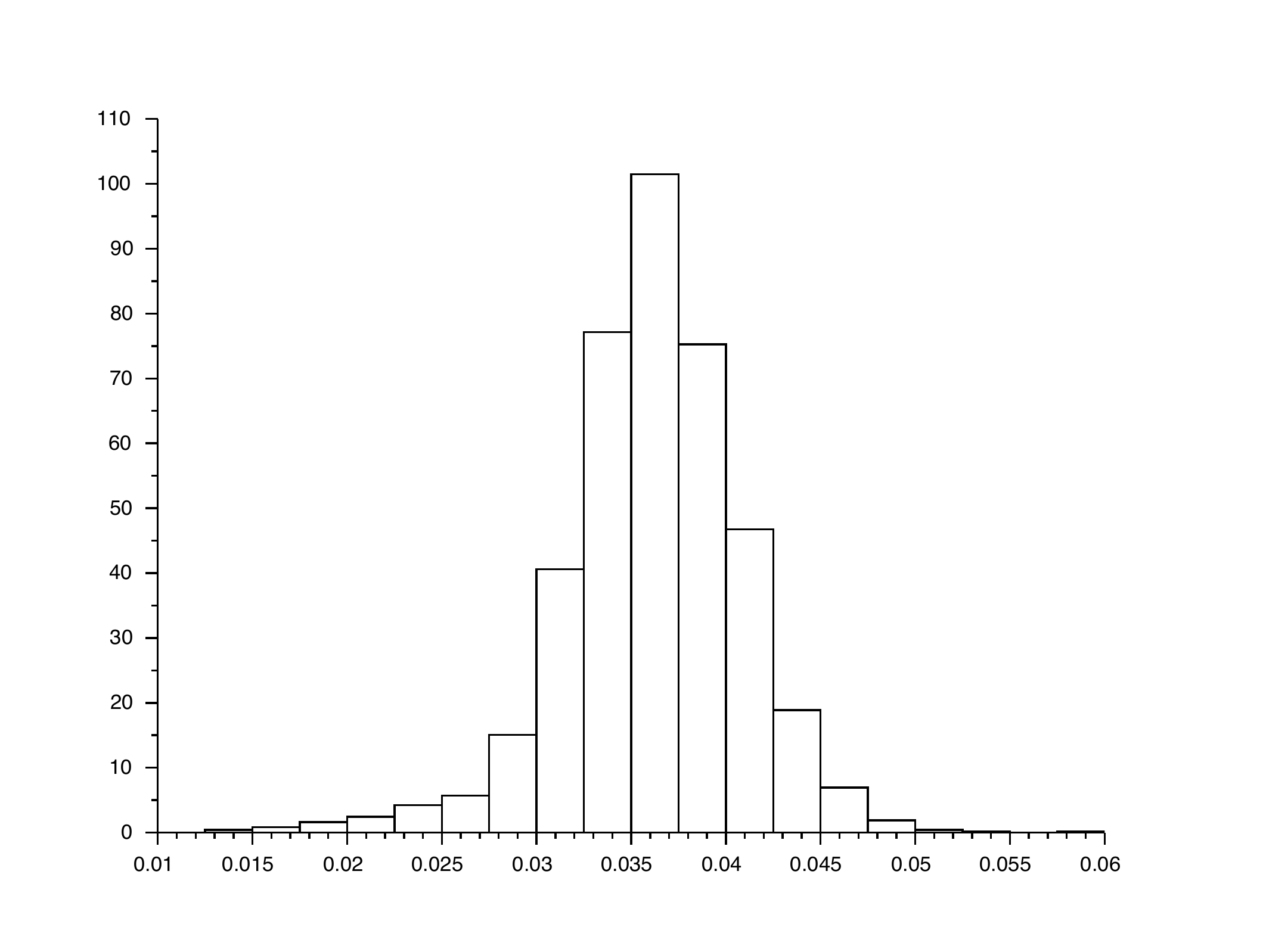}}
\subfigure[Generation 8]{\includegraphics[width=.9\linewidth]{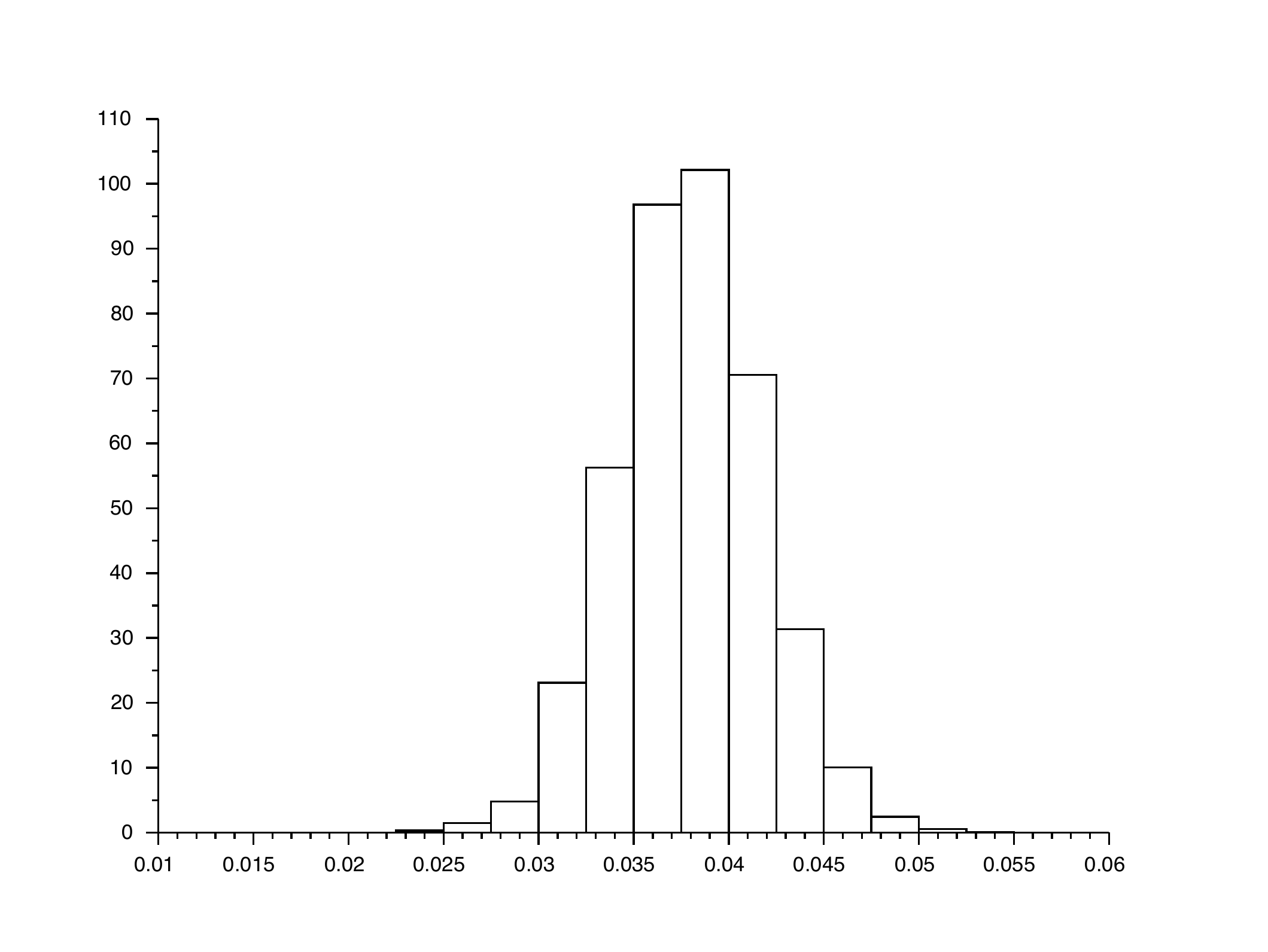}}
\caption{Histogram of growth rates for cells in generations 2 (a), 5 (b) and 8 (c), Stewart data set.}
\label{fig:hist-stewart}
\end{figure}

We then investigated the stationarity of the growth rate in the  data.
The two datasets correspond  to different experimental procedures, therefore creating potential differences in the initial physiological state of the cellls. In \cite{stewart}, the initial cells were picked at random from a population growing in a liquid medium and then plated on a solid medium, where it grew and divided to form  microcolonies. The cells undergo a plating stress when placed on the solid medium, which is well known  by biologists, see e.g. \cite{rolfe}  and  \cite{cuny}. This leads to a transient phase of reduced growth rates in the first generations, see Figures~\ref{fig:boxp-stewart} and \ref{fig:hist-stewart}. 
%
 In \cite{lydia}, on the contrary, the first generations of cells were removed, so that only a steady state is observed, see Figure~\ref{fig:boxp-wang} which is the counterpart of Figure~\ref{fig:boxp-stewart} and presents boxplots of the growth rates of cells for Wang's data for generations 2, 3, 4, 5, 10, 20, 30, 40, 50, 100 and 200.
For Wang's data set, one can be a bit more precise regarding stationarity for the cumulated old pole lineage.
We implemented the following procedure (on old pole cells only):
\begin{figure}[H]
\centering
\includegraphics[trim={3ex 10ex 3ex 12ex},width=\linewidth,clip]{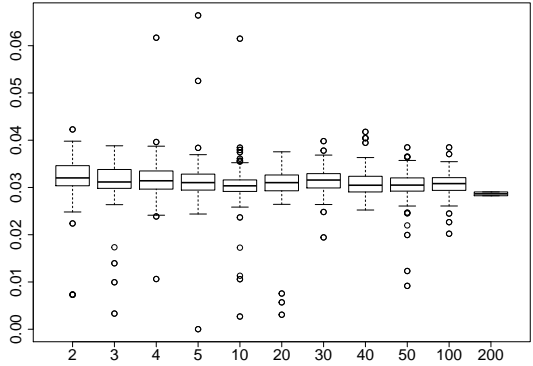}\\
\caption{Box plots of growth rates for cells in generations 2, 3, 4, 5, 10, 20, 30, 40, 50, 100 and 200, Wang data set. Outliers (growth rate negative or larger than 0.08) are excluded for clarity.}
\label{fig:boxp-wang}
\end{figure}
\begin{enumerate}
\item For each tree
\begin{enumerate}
\item    The residuals of an ARMA(1,1) model are computed. 
\item    These residuals are split first half / second half
\item    A Kolmogorov test (R command {\tt ks.test}) is used for comparison of distributions of the subseries.
\end{enumerate}
\item    We plot in Figure~\ref{fgkolm} an histogram of the p-values.
\end{enumerate}
\begin{figure}[H]
\begin{center}
\centerline{\includegraphics[trim={4ex 10ex 6ex 12ex},width=\linewidth]{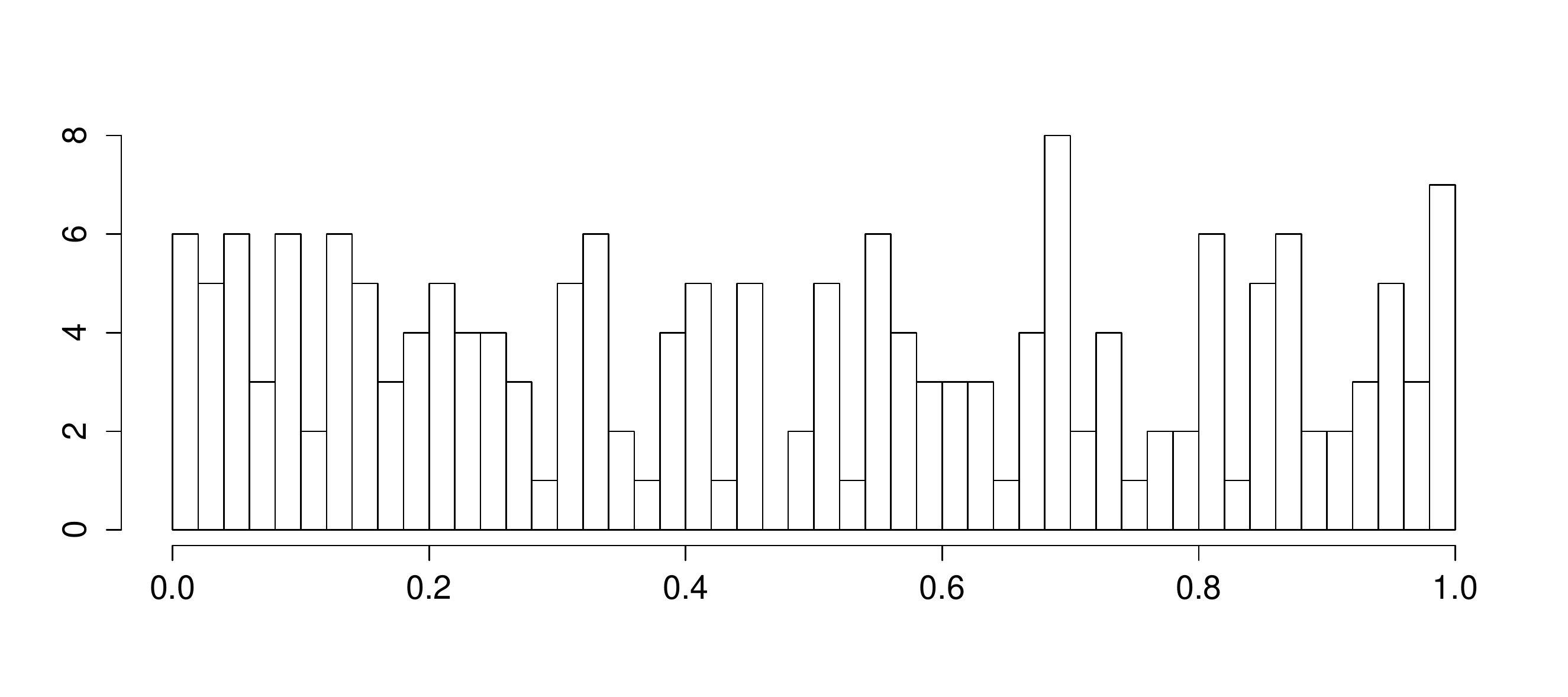}}
\caption{P-values for the Kolmogorov test of stationarity, Wang data set.}\label{fgkolm}
\end{center}
\end{figure}
%
Lets us explain our motivation for the first step. In order to 
use the right  threshold in the Kolmogorov test, we need in theory the data to be independent.
Assuming that the growth rate is an AR(1)
process, and that the data are noisy observations of the growth rate, we get indeed an ARMA(1,1)
process. Steps 2 and 3 are standard. Concerning step~4, under $H_0$ (stationarity),
the p-values are uniformly distributed, and else, they are more concentrated around 0. 
 \newline We see on Figure~\ref{fgkolm} a uniform distribution of the p-values, which is characteristic
of the non-significance of the hypothesis of different distributions.
We obtain the same conclusion if we replace the Kolmogorov test with a Student test 
(change {\tt ks.test} into {\tt t.test}).
%
\section{Conclusion}
\label{sec:ccl}
 In these two data sets we made efforts to take into account the tree structure of the data.
We tried  different statistical procedures that can be summed up as follows.
\paragraph{Wang data.} Because of the simple structure of this data set, each tree is here just the grey subtree in Figure\,\ref{fig:treetype}. We have tried dynamical models in which the growth rate of a cell
may have a multi-generation memory, with coefficients possibly dependent on the tree
(mixed effects). We did not find a significant improvement over the simplest model
where the rate of a cell depends only on the one of its mother,
and that of the grand mother has no significant influence. We found that
\begin{enumerate}
\item the old pole cell growth rate is  significantly more correlated to its mother than the 
new pole cell growth rate;
\item the mean old pole cell growth rate is significantly smaller 
than the mean new pole cell growth rate;
\item the stationarity cannot be rejected. 
\end{enumerate}
~
\paragraph{Stewart data.} The tree structure induces dependency in the data which we have take into account 
in our testing procedures. It is established in the literature that old pole and new pole cells have significantly different growth rates on this data set. In addition, we found the following. \begin{enumerate}
\item There is no stationarity of the growth rate across generations. 
This means that the initial stress of the
experiment has not the time to vanish during only the first 9 generations.
\item
The most relevant factor is the the number of generations since the last change of pole type, and not the whole sequence of types along the lineage of a given cell.
For example, cell 17 (NNO) in figure~\ref{fig:treetype} has a similar growth rate as 
cells 21 and 29 (ONO), or NONOONN (300, 428) as NNONN (68, 100).
\end{enumerate}
 To conclude, in both data sets, we recover a statistically significant difference between the growth rate of sister cells. Therefore, asymmetry is present in the division of the {\it E. coli}, even after hundreds of generations.
 \newline The apparent conflict between both data sets may simply come from observations at different phases: Stewart’s data are still in a transient phase whereas Wang’s data are stationary.
From this point of view, the two data sets are not contradictory.
To our best knowledge, there is no available data set of {\it E. coli} division with both
transient and steady states.
It would be interesting to design an experiment where both the transient and the stationary phase could be observed on the same colonies.

\section*{Acknowledgements}

The research of B. de Saporta and  N. Krell  is partly supported by the ANR, Agence Nationale de la Recherche, grant PIECE
12-JS01-0006-01. 

{The authors gratefully thank E. Stewart for numerous insightful discussions, careful reading of a preliminary version of the paper, and for giving permission to publish the data.}


\end{multicols}


\begin{thebibliography}{}

\bibitem[Cuny et~al., 2007]{cuny}
Cuny, C., Lesbats, M., and Dukan, S. (2007).
\newblock Induction of a global stress response during the first step of
  escherichia coli plate growth.
\newblock {\em Appl Environ Microbiol}, 73(3):885–889.

\bibitem[de~Saporta et~al., 2011]{SGM11}
de~Saporta, B., G{\'e}gout-Petit, A., and Marsalle, L. (2011).
\newblock Parameters estimation for asymmetric bifurcating autoregressive
  processes with missing data.
\newblock {\em Electron. J. Stat.}, 5:1313--1353.

\bibitem[de~Saporta et~al., 2012]{SGM12}
de~Saporta, B., G{\'e}gout-Petit, A., and Marsalle, L. (2012).
\newblock Asymmetry tests for bifurcating auto-regressive processes with
  missing data.
\newblock {\em Statist. Probab. Lett.}, 82(7):1439--1444.

\bibitem[de~Saporta et~al., 2014]{SGM14}
de~Saporta, B., G{\'e}gout-Petit, A., and Marsalle, L. (2014).
\newblock Statistical study of asymmetry in cell lineage data.
\newblock {\em Comput. Statist. Data Anal.}, 69:15--39.

\bibitem[Doumic et~al., 2015]{hof}
Doumic, M., Hoffmann, M., Krell, N., and Robert, L. (2015).
\newblock Statistical estimation of a growth-fragmentation model observed on a
  genealogical tree.
\newblock {\em Bernoulli}, 21(3):1760--1799.

\bibitem[Guyon, 2007]{Guy07}
Guyon, J. (2007).
\newblock Limit theorems for bifurcating {M}arkov chains. {A}pplication to the
  detection of cellular aging.
\newblock {\em Ann. Appl. Probab.}, 17(5-6):1538--1569.

\bibitem[Guyon et~al., 2005]{Guy05}
Guyon, J., Bize, A., Paul, G., Stewart, E., Delmas, J.-F., and Tadd{\'e}i, F.
  (2005).
\newblock Statistical study of cellular aging.
\newblock In {\em C{EMRACS} 2004---mathematics and applications to biology and
  medicine}, volume~14 of {\em ESAIM Proc.}, pages 100--114 (electronic). EDP
  Sci., Les Ulis.

\bibitem[Robert et~al., 2014]{hof2}
Robert, L., Hoffmann, M., Krell, N., Aymerichand, S., Robert, J., and Doumic,
  M. (2014).
\newblock Division in {\it escherichia coli} is triggered by a size-sensing
  rather than a timing mechanism.
\newblock {\em BMC Biology}, 12(17).

\bibitem[Rolfe et~al., 2012]{rolfe}
Rolfe, M.~D., Rice, C.~J., Lucchini, S., Pin, C., Thompson, A., Cameron, A.
  D.~S., Alston, M., Stringer, M.~F., Betts, R.~P., Baranyi, J., Peck, M.~W.,
  and Hinton, J. C.~D. (2012).
\newblock Lag phase is a distinct growth phase that prepares bacteria for
  exponential growth and involves transient metal accumulation.
\newblock {\em J Bacteriol.}, 194(3):686--701.

\bibitem[Stewart et~al., 2005]{stewart}
Stewart, E.~J., Madden, R., Paul, G., and Taddei, F. (2005).
\newblock Aging and death in an organism that reproduces by morphologically
  symmetric division.
\newblock {\em PLoS Biol}, 3(2):686--701.

\bibitem[Wang et~al., 2012]{lydia}
Wang, P., Robert, L., Pelletier, J., Dang, W.~L., Taddei, F., Wright, A., and
  Jun, S. (2012).
\newblock Robust growth of {\it escherichia coli}.
\newblock {\em Current Biology}, 20(12):1099 -- 1103.

\end{thebibliography}
\end{document}